\documentclass[5p,times,twocolumn]{elsarticle}
\usepackage{amsmath,amssymb,hyperref}
\usepackage[dvips]{epsfig}
\usepackage{graphicx}
\usepackage{xcolor}
\usepackage{bbm}

\definecolor{heidelbeer}{rgb}{0.5,0,0.5}

\newcommand{\bphi}{\boldsymbol\varphi}

\graphicspath{{./Figures/}}

\begin{document}

\begin{frontmatter}

\title{Optimization of Schwinger pair production in colliding laser pulses}

\author[hd]{F. Hebenstreit}
\ead{f.hebenstreit@thphys.uni-heidelberg.de}

\author[qbc]{F. Fillion-Gourdeau}
\ead{francois.fillion@emt.inrs.ca}

\address[hd]{Institut f\"{u}r Theoretische Physik, Universit\"{a}t Heidelberg, D-69120 Heidelberg, Germany}
\address[qbc]{Universit\'{e} du Qu\'{e}bec, INRS-\'{E}nergie, Mat\'{e}riaux et T\'{e}l\'{e}communications, Varennes, Qu\'{e}bec, Canada J3X 1S2}

\begin{abstract}
 Recent studies of Schwinger pair production have demonstrated that the asymptotic particle spectrum is extremely sensitive to the applied field profile.
 We extend the idea of the dynamically assisted Schwinger effect from single pulse profiles to more realistic field configurations to be generated in an all-optical experiment searching for pair creation.
 We use the quantum kinetic approach to study the particle production and employ a multi-start method, combined with optimal control theory, to determine a set of parameters for which the particle yield in the forward direction in momentum space is maximized. We argue that this strategy can be used to enhance the signal of pair production on a given detector in an experimental setup.
\end{abstract}


\end{frontmatter}

\section{Introduction}

The creation of electron-positron pairs from external electric fields (Schwinger effect) has been a long-standing prediction of quantum electrodynamics (QED) \cite{Sauter:1931zz,Heisenberg:1935qt,Schwinger:1951nm}.
The breakdown of the QED vacuum has not been observed yet because of the required electric field strength which is of the order of $E_S\sim10^{16}\, V/m$.
However, recent theoretical studies as well as technological advances have raised the hope that an experimental observation might become feasible in the near future \cite{Tajima:1900zz,Gordienko:2005zz}.

In recent years, investigations have demonstrated that the electron-positron spectrum is extremely sensitive to the applied electric field profile \cite{Hebenstreit:2009km,Dumlu:2010vv,Abdukerim:2013,Kohlfurst:2013ura,Akal:2014eua,Dumlu:2010ua,FillionGourdeau:2012qh,Blaschke:2013ip,Blinne:2013via}.
Most notably, it has been shown that the particle production can be significantly enhanced by using optimized or tailored field configurations.
In this respect, the dynamically assisted Schwinger effect was proposed as a mechanism to enhance non-perturbative particle production by orders of magnitude:
Superimposing a strong but low-frequency field with a weak but high-frequency field partially lifts the exponential suppression of the Schwinger effect due to dynamical pair creation \cite{Schutzhold:2008pz,Orthaber:2011cm,Fey:2011if,Jiang:2012if,Jansen:2013}.
On the other hand, it has also been shown that quantum interference can result in a drastic enhancement or decrease of the particle yield:
For a sequence of single pulses it was demonstrated that an order-of-magnitude variation of the particle yield occurs upon changing the interpulse time-lag while keeping all other parameters fixed \cite{Dumlu:2011rr,Akkermans:2011yn,Kohlfurst:2012rb,Li:2014}.
Most studies, however, have been based on very simple field configurations -- mostly superpositions of single electric pulses -- which are certainly not realistic in the sense of representing experimentally relevant fields.

A realistic field configuration, however, which is likely to be generated in an all-optical experiment searching for pair creation is supposed to be more complicated.
Theoretical models of such field configurations, which are in fact non-trivial solutions of Maxwell equations in vacuum, include standing-wave beam pulses or superpositions of e-dipole pulses \cite{Davis:1979zz,Overfelt:1991,Gonoskov:2012}.
Previous investigations of the asymptotic particle number (but not its spectrum) in these electromagnetic backgrounds have only been based on the locally-constant field approximation \cite{Bulanov:2004de,Bulanov:2010ei,Gonoskov:2013ada}.

To a first approximation, the electromagnetic field in the focal spot of these field configurations can also be approximated by a spatially homogeneous but time-dependent electric field which points in a given direction. This is based on the fact that the spatial scale for particle production, which is set by the Compton wavelength, is orders of magnitude smaller than typical scales of optical lasers.
The typical time profile of electric fields, which have been employed in the study of the asymptotic particle spectrum, are then given by an envelope with subcycle structure, including a carrier phase and/or a chirp \cite{Hebenstreit:2009km,Dumlu:2010vv,Abdukerim:2013,Kohlfurst:2013ura,Akal:2014eua}.

In this publication, we investigate the possibility of generating an optimized field configurations from a superposition of two colliding laser pulses of different amplitude and frequency.
Accordingly, we extend the idea of the dynamically assisted Schwinger effect from single pulse profiles to more realistic field configurations.
To this end, we employ the quantum kinetic formalism together with an optimization method to investigate the electron-positron spectrum \cite{Kohlfurst:2012rb}.
In fact, similar optimization techniques have proven succesful in the related field of atomic, molecular and optical physics (AMO), i.e. in the optimization of the higher harmonic yield or the generation of single attosecond pulses \cite{Chu:2001,Christov:2001,BenHajYedder:2004,Balogh:2014}.

We have to face an additional experimental challenge regarding the optimization of the particle yield in, for instance, e-dipole pulses: 
Corresponding experiments will require large parabolic mirrors to focus the laser beam and these mirrors are supposed to cover a large part of the solid angle.
Accordingly, the particles should ideally be emitted in some specific direction, where it is still possible to put a particle detector. 
Moreover, electron spectrometers are usually sensitive in a certain range of momenta and are characterized by a finite momentum resolution.
In this sense, we will focus here on the maximization of produced particles along the electric field direction within a specific range of finite momenta.

This publication is organized in the following way:
Section \ref{sec:theory} is devoted to reviewing the quantum kinetic equations for Schwinger pair production and discussing the optimization method for our problem.
In section \ref{sec:model}, we discuss our model and describe the superposition of two laser pulses with sub-cycle structure.  
We present our results regarding the dynamically assisted Schwinger effect for two colliding laser pulses and the optimal set of parameters, for which the particle yield in the forward direction in momentum space is maximized, in section \ref{sec:results}. 
We conclude and give an outlook in section \ref{sec:conclusion}.

\section{Quantum kinetics and optimization method}
\label{sec:theory}

There are various equivalent ways to calculate the asymptotic particle spectrum for Schwinger pair production in a time-dependent electric field $\mathbf{E}(t)=(0,0,E(t))$, represented by a vector potential $\mathbf{A}(t)=(0,0,A(t))$ such that $E(t)=-\dot{A}(t)$.
In this investigation, we employ the quantum kinetic formalism within which all spectral information is encoded in the distribution function $F(\mathbf{q},t)$, where $\mathbf{q}$ denotes the canonical momentum whose component parallel and perpendicular to the electric field are denoted by $q_\parallel$ and $\mathbf{q}_\perp$, respectively \cite{Kluger:1992gb,Schmidt:1998vi}.
The corresponding kinetic momenta are then given by $p_\parallel(t)=q_\parallel-eA(t)$ and $\mathbf{p}_\perp=\mathbf{q}_\perp$.
As there is no unique definition of a particle number in an interacting quantum field theory, $F(\mathbf{q},t)$ corresponds to the momentum spectrum of physical particles only at asymptotic times $\pm T$ when all interactions are switched off \cite{Dabrowski:2014ica}.

Assuming that the created particles do not act back on the prescribed electric field, the distribution function fulfills the system of differential equations \cite{Bloch:1999eu}:
\begin{subequations}
 \label{eq:qke}
 \begin{align}
 \dot{F}(\mathbf{q},t) &=  W(\mathbf{q},t)G(\mathbf{q},t) \ ,  \\
 \dot{G}(\mathbf{q},t) &=  W(\mathbf{q},t)[1-F(\mathbf{q},t)] - 2\omega(\mathbf{q},t) H(\mathbf{q},t) \ ,  \\
 \dot{H}(\mathbf{q},t) &=  2\omega(\mathbf{q},t) G(\mathbf{q},t) \ , 
 \end{align}
\end{subequations}
with initial conditions $F(\mathbf{q},-T)=G(\mathbf{q},-T)=H(\mathbf{q},-T)=0$.
Here, we introduced:
\begin{equation}
  W(\mathbf{q},t) = \frac{eE(t)\epsilon_\perp}{\omega^2(\mathbf{q},t)} \ ,
\end{equation}
with $\epsilon_\perp^2=m^2+\mathbf{q}_\perp^2$ and $\omega^2(\mathbf{q},t)=\epsilon_\perp^2+p_\parallel^2(t)$. For a given external field configuration, these equations are solved numerically by standard ordinary differential equation solver.
The number of created particles in a specific momentum space volume $\Omega$ is then defined by:
\begin{equation}
 \label{eq:num}
 n[e^+e^-;\Omega]=\int_\Omega{\frac{d^3q}{(2\pi)^3}F(\mathbf{q},T)} \ .
\end{equation}

In order to formulate the optimization problem, we define the cost functional according to \cite{Kohlfurst:2012rb}:
\begin{equation}
 J[F,A]=-n[e^+e^-;\Omega] \ ,
\end{equation}
where $\Omega$ denotes the momentum space volume in which the asymptotic density of created particles is to be maximized.
The optimization procedure then changes the vector potential $A(t)$ such that the cost functional becomes minimized. Thus, the optimization problem can be written as
\begin{equation}
\label{eq:opt}
\tilde{J} = 
\min_{\{\varphi_i|i=1,\cdots,n\}} J[F[A],A] \ ,
\end{equation}
where $\tilde{J}$ is the global minimum in parameter space and where it is assumed that the vector potential is parametrized by the set $(\varphi_{i})_{i=1,\cdots,n}$, i.~e. written as $A(t,\boldsymbol\varphi)$. In the following investigation, we will prescribe the functional form of this vector potential but still allow  two carrier phases as well as the time-lag between the two pulses to be optimized, corresponding to $n=3$ (this will be discussed in more details in the next section). 

To solve the optimization problem \eqref{eq:opt}, we employ a multi-start method. 
This class of numerical method consists of two phases: 
After generating random solutions $\boldsymbol\varphi_{0}$, these trial solutions are further improved in a second step. In this work, the random solutions are generated using a simple constant probability distribution in parameter space. It has been shown that such an approach converges towards the global minimum as the number of sample points goes to infinity \cite{marti:2003}. Moreover, the multi-start methods can be parallelized easily, resulting in an efficient numerical method to search our relatively low dimensional parameter space. On the other hand, the improvement of solutions is performed by optimal control theory based on the Broyden-Fletcher-Goldfarb-Shanno (BFGS) algorithm.

In fact, the optimization problem is a constrained one as $F(\mathbf{q},T)$ is supposed to be the solution of the equations of motion \eqref{eq:qke}.
This constraint can be taken into account by introducing Lagrange multipliers $\lambda_{F}(\mathbf{q},t)$, $\lambda_{G}(\mathbf{q},t)$, $\lambda_{H}(\mathbf{q},t)$ which have to fulfill the adjoint equations:
\begin{subequations}
 \label{eq:adj}
 \begin{align}
 \dot{\lambda}_F(\mathbf{q},t) &=  W(\mathbf{q},t)\lambda_G(\mathbf{q},t) \ ,  \\
 \dot{\lambda}_G(\mathbf{q},t) &=  -W(\mathbf{q},t)\lambda_F(\mathbf{q},t) - 2\omega(\mathbf{q},t) \lambda_H(\mathbf{q},t) \ ,  \\
 \dot{\lambda}_H(\mathbf{q},t) &=  2\omega(\mathbf{q},t) \lambda_G(\mathbf{q},t) \ , 
 \end{align}
\end{subequations}
with final conditions $\lambda_G(\mathbf{q},T)=\lambda_H(\mathbf{q},T)=0$ and:
\begin{equation}
 \lambda_F(\mathbf{q},T)= \begin{cases}  1 &\mathbf{q}\in\Omega \\ 0 &\mathbf{q}\notin\Omega \end{cases}
\end{equation}
In the following, these adjoint equations will be numerically solved by standard methods.
Given that $F$, $G$, $H$ is the unique solution of \eqref{eq:qke} and $\lambda_F$, $\lambda_G$, $\lambda_H$ fulfill the adjoint equations \eqref{eq:adj}, the formerly constrained optimization problem can be restated as an unconstrained optimization problem of the reduced cost functional:
\begin{equation}
 \hat{J}(\bphi)=J[F(A(\boldsymbol\varphi)),A(\bphi)] \ .
\end{equation}
Its gradient $\nabla_{\bphi}\hat{J}\in\mathbbm{R}^n$, which is required for the optimization procedure to be employed, is then given by:
\begin{alignat}{2}
 \label{eq:grad} 
 \nabla_{\bphi}\hat{J}(\bphi) = e&\int\limits_{-\infty}^\infty\frac{d^3q}{(2\pi)^3}\int\limits_{-T}^T{dt}\left[\frac{2p_\parallel}{\omega}\left(\mu_HG-\mu_GH\right)\nabla_\mathbf{\bphi} A \right. \nonumber \\
                                &+\left.\left(\mu_GF-\mu_FG-\mu_G\right)\left(\frac{\epsilon_\perp}{\omega^2}\nabla_{\bphi}E+\frac{E\epsilon_\perp p_\parallel }{\omega^4}\nabla_{\bphi}A\right)\right] \  .
\end{alignat}

For a given trial configuration $\bphi_0$, a local minimizer of the reduced cost functional is iteratively found via:
\begin{equation}
 \bphi_{k+1} = \bphi_{k} + \alpha_k \mathbf{d}_k \ , 
\end{equation}
with $k\in\mathbbm{N}^0$.
In the current study, we use \eqref{eq:grad} to calculate the search direction $\mathbf{d}_k$ according to the BFGS algorithm.
Moreover, we perform an inexact line search to determine a viable step size $\alpha_k$ fulfilling the strong Wolfe conditions.
For further algorithmic details we refer to \cite{Nocedal:1999}. 


\section{Temporal field profile}
\label{sec:model}

There have been various investigations of electron-positron production in standing-wave beam pulses or superpositions of e-dipole pulses.
The appealing feature of these field configurations is that they represent actual solutions of Maxwell equations in vacuum.
A major drawback, however, is that particle production could only be investigated based on the locally-constant field approximation \cite{Bulanov:2004de,Bulanov:2010ei,Gonoskov:2013ada}.

In this publication, we take another viewpoint and regard the electromagnetic field as a spatially homogeneous, time-dependent electric field which points into a given direction.
Accordingly, the quantum kinetic formalism as presented in the previous section applies.
The assumption of spatial homogeneity is based on the fact that the spatial scale for particle production, which is set by the Compton wavelength, is orders of magnitude smaller than typical scales of optical lasers. This can be achieved for an e-dipole field in the vicinity of the focal point. 
In such a case, it can be shown that the magnetic field vanishes at the focus such that the time-dependence is just determined by the driving longitudinal field \cite{Gonoskov:2012}.

It has to be emphasized that the approximation of spatial homogeneity is strictly valid only at this particular position, whereas electric and magnetic field components are still existent in other regions. In principle, the magnetic field and the corresponding spatially dependent vector potential should be taken into account. However, the particle production in these regions is supposed to be suppressed due to the weaker electric field strength as compared to the field in the vicinity of the focal spot. As the latter gives the main contribution to the pair production rate, we completely neglect the spatial inhomogeneity.
In fact, an ab initio investigation of the full problem is computationally not feasible yet.

To be specific, we consider the superposition of two oscillating electric fields with Gaussian envelope, which represent the field at the focus of two colliding laser pulses:
\begin{equation}
 E_j(t) = E_j\cos(\omega_jt - \phi_j)e^{-\tfrac{t^2}{2\tau_j^2}}\ ,
\end{equation}
with $j=\{1,2\}$, such that the total electric field is given by:
\begin{equation}
 \label{eq:efld}
 E(t) = E_1(t) + E_2(t-T) \ .
\end{equation}
Here, $E_j$ are the peak field strengths, $\omega_j$ are the laser frequencies, $\tau_j$ define the total pulse lengths, $\phi_j$ are carrier phases and $T$ denotes the time-lag between the two pulses.
We note that the combination $\sigma_j\equiv\tau_j\omega_j$ determines the number of cycles within the pulse.
From an experimental point of view, the parameters $\{E_j,\omega_j,\tau_j\}$ are restricted by the actual facilities whereas the carrier phases and time-lag are still free parameters.
The corresponding vector potentials are then given by:
\begin{equation}
 A_j(t) = -\sqrt{\tfrac{\pi}{2}}E_j\tau_je^{-\tfrac{(\tau_j\omega_j)^2}{2}}\operatorname{Re}\left[e^{i\phi_j}\operatorname{erf}\left(\tfrac{t+i\tau_j^2\omega_j}{\sqrt{2}\tau_j}\right)\right] \ ,
\end{equation}
with $j=\{1,2\}$, such that:
\begin{equation}
 A(t) = A_1(t) + A_2(t-T) \ .
\end{equation}

Regarding the optimization problem, we will consider the carrier phases and the time-lag as parameters to be optimized, i.~e. $\bphi=(\phi_1,\phi_2,T)$.
The derivatives of the electric field with respect to the parameters are given by:
\begin{subequations}
 \begin{align}
  \partial_{\phi_1}E(t) &= E_1\sin(\omega_1t-\phi_1)e^{-\tfrac{t^2}{2\tau_1^2}} \ , \\
  \partial_{\phi_2}E(t) &= E_2\sin(\omega_2s-\phi_2)e^{-\tfrac{s^2}{2\tau_2^2}} \ , \\
  \partial_{T}E(t) &= E_2\left[\omega_2\sin(\omega_2s-\phi_2) + \frac{s}{\tau_2^2}\cos(\omega_2s-\phi_2)\right]e^{-\tfrac{s^2}{2\tau_2^2}} \ ,
 \end{align}
\end{subequations}
with $s=t-T$,
whereas the derivatives of the vector potential read:
\begin{subequations}
 \begin{align}
  \partial_{\phi_1}A(t) & = \sqrt{\tfrac{\pi}{2}}E_1\tau_1e^{-\tfrac{(\tau_1\omega_1)^2}{2}}\operatorname{Im}\left[e^{i\phi_1}\operatorname{erf}\left(\tfrac{t+i\tau_1^2\omega_1}{\sqrt{2}\tau_1}\right)\right]\ , \\
  \partial_{\phi_2}A(t) & = \sqrt{\tfrac{\pi}{2}}E_2\tau_2e^{-\tfrac{(\tau_2\omega_2)^2}{2}}\operatorname{Im}\left[e^{i\phi_2}\operatorname{erf}\left(\tfrac{s+i\tau_2^2\omega_2}{\sqrt{2}\tau_2}\right)\right]\ , \\
  \partial_{T}A(t) & = E_2\cos(\omega_2s - \phi_2)e^{-\tfrac{s^2}{2\tau_2^2}} \ .
 \end{align}
\end{subequations}
As a reminder, the field gradients $\nabla_{\bphi} A$ and $\nabla_{\bphi} E$ are required for calculating the gradient of the reduced cost functional \eqref{eq:grad}.

\section{Results}
\label{sec:results}

In the following we present our results:
First, we discuss the dynamically assisted Schwinger mechanism for a 'naive' superposition of two pulses with $\phi_1=\phi_2=T=0$.
In this part, we investigate the possible enhancement of the particle yield by fixing the field strength and varying $\omega_2$ and $\tau_2$ such that $\sigma_2=\tau_2\omega_2=\text{const}$. 
Subsequently, we consider a specific set of parameters (field strengths, frequencies, pulse lengths) and study the optimization with respect to the parameters $\bphi=\{\phi_1,\phi_2,T\}$.
According to the previous discussion we show in this part that the particle yield in specific ranges of momentum space can be further enhanced by an optimal superposition of the two pulses while keeping all remaining parameters fixed.

\subsection{Dynamically assisted Schwinger mechanism}

We consider the superposition of two oscillating electric fields with Gaussian envelope \eqref{eq:efld} with $\bphi_0=(0,0,0)$.
Most notably, the first pulse is in the adiabatic regime whereas the second pulse is in the anti-adiabatic regime.
These different regimes are discriminated by the Keldysh parameter \cite{Brezin:1970xf}:\footnote{The Keldysh parameter was originally defined only for a monochromatic electric field. For non-monochromatic field configurations one usually takes the most dominant frequency scale.} 
\begin{equation}
 \gamma_i=\frac{m\omega_i}{eE_i}=\frac{E_S}{E_i}\frac{\omega_i}{m} \ .
\end{equation}
For the adiabatic pulse ($\gamma_1\ll1$) we take a $\sigma_1=5$-cycle soft X-ray pulse with peak field strength of $E_1=0.1E_S$, frequency $\omega_1=m/40$ and pulse length parameter $\tau_1=200/m$.
The anti-adiabatic pulse ($\gamma_2\gg1$) is taken to be a $\sigma_2=20$-cycle hard X-ray pulse (thus $\tau_{2} = 20/\omega_{2}$) with peak field strength $E_2=0.01E_S$.

\begin{figure}[t]
 \centering
 \includegraphics[width=0.95\columnwidth]{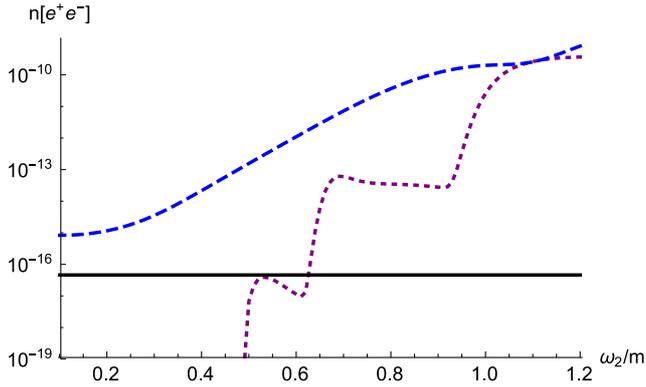}
 \caption{Logarithmic plot of the asymptotic particle numbers $n_1[e^+e^-]$ (solid, black), $n_2[e^+e^-]$ (dotted, purple) and $n_{1+2}[e^+e^-]$ (dashed, blue) for $E_1=0.1E_S$, $\omega_1=m/40$, $\tau_1=200/m$ and $E_2=0.01E_S$ as function of $\omega_2/m$.}
 \label{fig1}
\end{figure}

In Fig.~\ref{fig1} we show the asymptotic particle densities $n_{1+2}[e^+e^-]$, generated by the electric field $E_1(t)+E_2(t)$, as well as $n_j[e^+e^-]$, with $j=\{1,2\}$, as function of $\omega_2/m$.
Notably, we observe a non-monotonic behavior of $n_2[e^+e^-]$ which is caused by the smeared multiphoton absorption thresholds, resulting in a sharp rise when the condition $N\omega_{2} \simeq 2m$ is fulfilled, with $N \in \mathbb{N}$ being the number of photons \cite{Kohlfurst:2013ura,Akal:2014eua,Heinzl:2010vg,Mocken:2010,Hebenstreit:2010cc}.
We emphasize that the value of the distribution function resulting in $n_2[e^+e^-]$ is too small to be resolved by our numerics below the threshold $\omega_2\lesssim m/2$.
In this regime only the adiabatic pulse gives a sizeable particle production so that we have $n_2[e^+e^-]\ll n_1[e^+e^-]$. 
On the other hand, above the threshold $\omega_2\gtrsim 2m/3$, we find $n_2[e^+e^-]\gg n_1[e^+e^-]$ so that particle production by the anti-adiabatic pulse is much more efficient.

The combined particle density $n_{1+2}[e^+e^-]$, on the other hand, shows a monotonic behavior as a function of $\omega_2$, indicating that the distinct threshold structure is washed out. 
Moreover, its magnitude can be orders of magnitude larger than the particle densities $n_1[e^+e^-]$ and $n_2[e^+e^-]$, respectively. 
This shows that the dynamically assisted Schwinger mechanism can result in an enhancement as large as a few orders of magnitude for the field configuration \eqref{eq:efld}.

\begin{figure}[t]
 \centering
 \includegraphics[width=0.95\columnwidth]{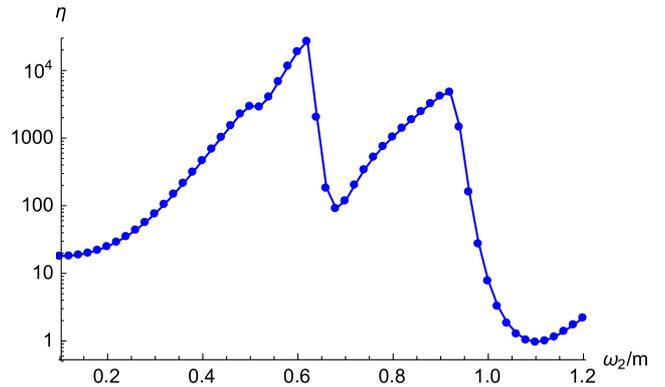}
 \caption{Logarithmic plot of the relative enhancement $\eta$ due to the dynamically assisted Schwinger mechanism as function of $\omega_2/m$. The other parameters are as in Fig.~\ref{fig1}.}
 \label{fig2}
\end{figure}

A simple measure for the enhancement is given by \cite{Orthaber:2011cm}: 
\begin{equation}
 \eta=\frac{n_{1+2}[e^+e^-]}{n_1[e^+e^-]+n_2[e^+e^-]} \ . 
\end{equation}
This relative enhancement, which results from the non-linear behavior of the particle production in combined electric fields, shows its typical behavior as a function of $\omega_2$.
For large values of $\omega_2$ we expect $\eta\simeq\mathcal{O}(1)$ since the multiphoton pair creation completely dominates the non-perturbative Schwinger mechanism. 
For small values of $\omega_2$, on the other hand, the relative enhancement is also expected to be $\eta\simeq\mathcal{O}(1-10)$, with the residual enhancement related to the higher peak field strength $E_1+E_2$.
At intermediate regimes, however, the relative enhancement can reach $\eta\simeq\mathcal{O}(10^4)$ for the chosen parameters, as shown in Fig.~\ref{fig2}.
The maxima in $\eta$ correspond to the frequencies directly below the smeared multiphoton absorption thresholds. 
Accordingly, they are peaked around $\omega_2\simeq0.6m$ as well as $\omega_2\simeq0.9m$.

\begin{figure}[b]
 \centering
 \includegraphics[width=0.95\columnwidth]{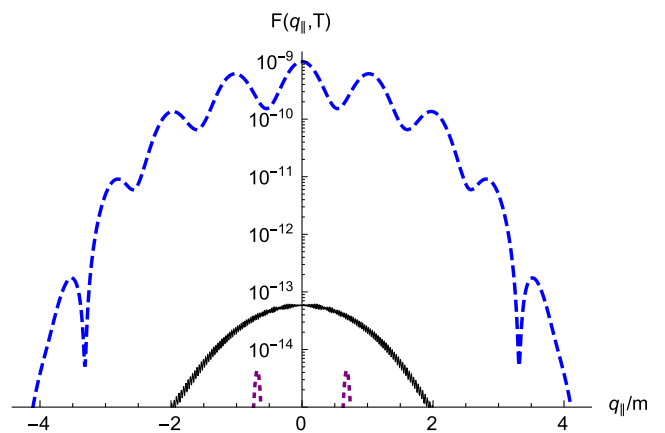}
 \caption{Logarithmic plot of the asymptotic momentum spectra $F_1(q_\parallel,T)$ (solid, black), $F_2(q_\parallel,T)$ (dotted, purple) and $F_{1+2}(q_\parallel,T)$ (dashed, blue) for $|\mathbf{q}_\perp|=0$ and $\omega_2=0.6m$. The other parameters are as in Fig.~\ref{fig1}.}
 \label{fig3}
\end{figure}

One strength of the quantum kinetic approach is that one has direct access to the momentum spectrum $F(\mathbf{q},T)$ of created particles.
In Fig.~\ref{fig3} we compare the distribution functions at $|\mathbf{q}_\perp|=0$ for $\omega_2=0.6m$, i.e. the value where the maximum relative enhancement was found.\footnote{We normalize the vector potential $A(T)=0$ so that $p_\parallel(T)=q_\parallel$.}
We note that the distribution function show the characteristic exponential falloff behavior for $|\mathbf{q}_\perp|>0$, with the scale set by the mass $m$.
Even though a destinctive cellular structure can be observed \cite{Blaschke:2013ip}, we do not study the momentum spectrum as function of $|\mathbf{q}_\perp|$ in more detail.

The distribution function $F_1(q_\parallel,T)$ shows the typical behavior of an envelope with small oscillations on top of it, with the oscillation scale $\Delta q_\parallel=\omega_1$ \cite{Hebenstreit:2009km}.
The distribution function $F_2(q_\parallel,T)$, on the other hand, shows the typical multiphoton absorption peaks \cite{Kohlfurst:2013ura}.
Most notably, there is no sign of the absorption peaks left in $F_{1+2}(q_\parallel,T)$.
Moreover, the oscillation scale of the distribution function does not depend of $\omega_1$ anymore but on the numerically determined parameter $\Delta q_\parallel=m$.
In fact, this parameter shows a non-trivial dependence as a function of $\omega_2$.

It has to be emphasized that the central peak of the distribution function for the chosen field configuration is located at $q_\parallel=0$, whereas the first side peaks are found at $q_\parallel=\pm m$.
For even higher momenta, the distribution function decreases very quickly.
This means, however, that the largest fraction of the produced particles is non-relativistic and therefore supposed to stay comparatively long in the interaction zone and show further non-trivial dynamics.
However, in order to obtain clear signals from the dynamically assisted Schwinger mechanism, it would be favorable to produce a larger fraction of particles with relativistic energies such that they leave the focus without further interactions. 
In the following section we investigate the possibility of shifting a larger part of the produced particles to higher momenta $q_\parallel>m$ by using an optimization method.

\subsection{Optimized pair production in the forward direction}
As noted in the introduction, there are several experimental challenges regarding the optimization of the particle density in a realistic setup:
First, the particles should predominantly be emitted in some specific direction in which it is possible to put a particle detector.
Secondly, electron spectrometer are sensitive in a certain range of momenta and are characterized by a finite momentum resolution.
Finally, produced particles should leave the focal region without further interactions in order to obtain a distinct signal. 

In order to fulfill these requirements, we seek an optimal field configuration such that particle production around $|\mathbf{q}_\perp|\simeq0$, i.e. along the electric field direction, is maximized within a certain momentum range $q_\text{low}<q_\parallel<q_\text{high}$. 
Moreover, as we suppose that the parameters $\{E_j,\omega_j,\tau_j\}$ are restricted by the actual facilities, we only allow the carrier phases and time-lag to be optimized.

To be specific, we will consider $q_\text{low}=1.5m$ and $q_\text{high}=2.5m$ in the following so that the momentum space volume in which the asymptotic particle density should be maximized is given by:\footnote{The distribution function $F(\mathbf{q},T)$ is approximately constant as a function of $|\mathbf{q}_\perp|$ for variations of the order of $\Delta|\mathbf{q}_\perp|\equiv m\delta_\perp\ll m$. Since \eqref{eq:num} vanishes at $|\mathbf{q}_\perp|=0$, we have to consider a small but finite momentum range $\delta_\perp\ll 1$.}
\begin{equation}
 \Omega=\left\{\left.\mathbf{q}\,\right.|\,1.5m<q_\parallel<2.5m\ \wedge\ |\mathbf{q}_\perp|\leq m\delta_\perp\right\} \ .
\end{equation}
This choice is based on the assumption that the detector is put on the positive $z$--axis.
In a more realistic experimental setup, the same optimization procedure could be performed for a range $\Omega$ that corresponds to the solid angle covered by the detection apparatus. 

In the numerical calculation, we take again $E_1=0.1E_S$, $\omega_1=m/40$, $\tau_1=200/m$ as well as $E_2=0.01E_S$, $\omega_2=0.6m$, $\tau_2=20/\omega_2$. According to the multi-start method, we chose the initial values of the carrier phases and the time-lag to be uniformly distributed in the interval $\phi_{j,0}\in\{-\pi,\pi\}$ and $T_0\cdot m \in\{-10,10\}$, respectively. 
In this way, we are able to scan the whole parameter space for the most favorable maximum.

\begin{table}[t]
  \centering
  \begin{tabular}{r|r|r|c|c}
  $ \phi_{1,\text{opt}} \ $ & $\ \phi_{2,\text{opt}} \ $& $\ T_{\text{opt}}/\tau_2 \ $& $\ n[e^+e^-;\Omega]_\text{opt}/\delta_\perp^2 \ $ & $ \frac{n[e^+e^-;\Omega]_\text{opt}}{n[e^+e^-;\Omega]_{\bphi_0}} $ \\
  \hline
  $2.462\ $&$ 2.426\ $&$-0.302\ $ & $3.021\cdot10^{-12} $&$4.029$ \\
  $2.235\ $&$ 2.862\ $&$-0.482\ $ & $3.020\cdot10^{-12} $&$4.028$ \\
  $2.134\ $&$ 2.233\ $&$-0.704\ $ & $3.020\cdot10^{-12} $&$4.028$ \\
  \end{tabular}
  \caption{\label{tab1}
  Optimized parameters $\bphi_{\text{opt}}$ for the three largest local maxima $n[e^+e^-;\Omega]_\text{opt}$.
  The remaining parameters are $E_1=0.1E_S$, $\omega_1=m/40$, $\tau_1=200/m$ and $E_2=0.01E_S$, $\omega_2=0.6m$, $\tau_2=20/\omega_2$.} 
\end{table} 

\begin{figure}[b]
 \centering
 \includegraphics[width=0.95\columnwidth]{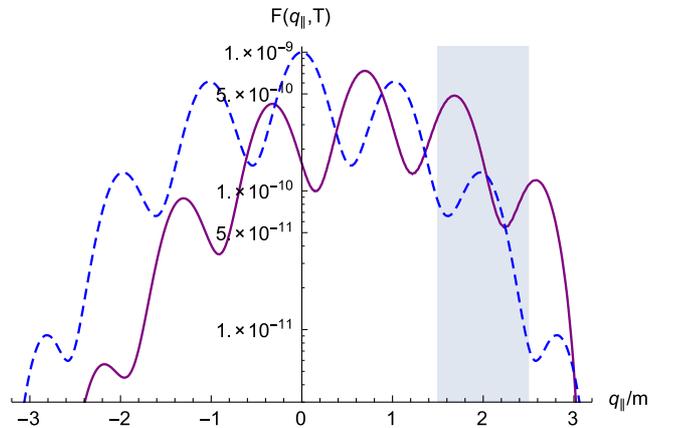}
 \caption{Asymptotic momentum spectra $F_\text{opt}(q_\parallel,T)$ (solid, purple) and $F_{\bphi_0}(q_\parallel,T)$ (dotted, blue) for $|\mathbf{q}_\perp|=0$.
 The shaded region corresponds to the momentum range $\Omega$. The other parameters are as in Tab.~\ref{tab1}.}
 \label{fig4}
\end{figure}

In Table~\ref{tab1} we present the three largest maxima $n[e^+e^-;\Omega]_\text{opt}$ which have been found by the optimization algorithm after convergence (requiring $100$ initial conditions).
These are compared with $n[e^+e^-;\Omega]_{\bphi_0}=7.497\cdot10^{-13}\delta_\perp^2$, which is obtained from the 'naive' superposition \eqref{eq:efld} with $\bphi_0=(0,0,0)$.
Obviously, the efficiency of the particle production in the selected momentum range $\Omega$ can be enhanced by a factor of $\simeq4$ by an optimal choice of parameters $\bphi_{\text{opt}}$.
On the other hand, we also emphasize that an inappropriate choice of parameters can result in an order of magnitude decrease of the particle production in $\Omega$.

In Fig.~\ref{fig4} we compare the optimized spectrum $F_\text{opt}(q_\parallel,T)$ with the asymptotic particle distribution $F_{\bphi_0}(q_\parallel,T)$.
Most notably, we observe that the optimization algorithm modifies the field configuration such that a larger amount of particles is shifted into the momentum range $\Omega$.
On the other hand, we also see that the particle distribution in other regions of momentum space decreases significantly.
As a consequence, we find for the total particle numbers $n[e^+e^-]_{\bphi_0}=1.095\cdot10^{-12}$ and $n[e^+e^-]_{\text{opt}}=7.902\cdot10^{-13}$, so that $n[e^+e^-]_{\bphi_0}>n[e^+e^-]_\text{opt}$.

\begin{figure}[t]
 \centering
 \includegraphics[width=0.95\columnwidth]{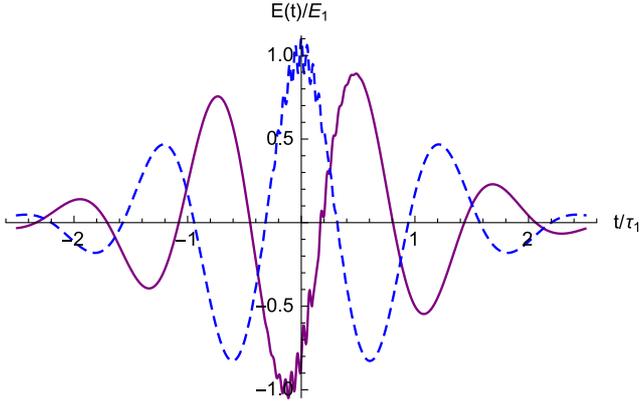}
 \caption{Comparison of the electric field $E(t)$ for $\bphi_\text{opt}=(2.462,2.426,-0.302)$ (solid, purple) with the naive superposition for $\bphi_0=(0,0,0)$ (dotted, blue). 
 The other parameters are as in Tab.~\ref{tab1}.}
 \label{fig5}
\end{figure}

The field configuration $E_\text{opt}(t)$, corresponding to the most favourable values of $\bphi_\text{opt}$, is shown in Fig.~\ref{fig5}.
Most notably, compared to the `naive' superposition, we find a rather large carrier phase $\phi_1\simeq2.4$ and small time-lag $T\simeq-0.3\tau$.
It has to be emphasized that the optimal values $\bphi_\text{opt}$ are hard to predict a priori as the Schwinger effects depends non-linearly on the electric field.
This can be understood most easily in the equivalent over-the-barrier scattering problem \cite{Dumlu:2010ua,Dumlu:2011rr}:
The complicated pattern of the distribution function results from quantum interference of pairs of complex conjugate turning points.
Accordingly, minor changes of $\bphi$ can have drastic effects as the actual positions of the turning points depend non-trivially on the field parameters and momentum $\mathbf{q}$.

We conclude that the optimal parameter choice strongly depends on the section of momentum space $\Omega$ which can be covered by particle detectors.
Having the laser parameters $\{E_j,\omega_j,\tau_j\}$ fixed, one can then indeed further increase the detection probability in $\Omega$ by just varying the carrier phases $\phi_j$ and the time-lag $T$.
This, however, goes at the expense of the production probability in other parts of momentum space.
Consequently, the notion of optimality is relative in the sense that it strongly depends on the momentum space section $\Omega$ under consideration.

\section{Conclusions \& Outlook}
\label{sec:conclusion}

We studied the dynamically assisted Schwinger effect for a superposition of two oscillating electric fields with Gaussian envelope.
We demonstrated that the combination of two laser pulses in the adiabatic and anti-adiabatic regime, respectively, can result an enhancement of the particle yield of the order of $\mathcal{O}(10^4)$.
It was shown, however, that the largest fraction of produced particles is created with vanishing kinetic momentum, suggesting that these particle are hard to directly detect.

Accordingly, we then focused on the possibility of shifting a larger amount of produced particles to higher momenta such that they leave the focal region without further interactions.
To this end, we assumed that characteristic laser parameters are fixed (peak field strength, frequency, pulse duration) whereas time lag and carrier phases are still tunable.
By employing an optimization method, we could determine a set of parameters such that the particle yield in a specific range of finite momenta was further increased by a factor of $\simeq4$.
It has to be pointed out, however, that the actual enhancement factor strongly depends on the momentum space section under consideration.
In the long run, such optimized field configurations could further facilitate the observation of the Schwinger effect in upcoming high-intensity laser experiments at the Extreme Light Infrastructure (ELI) or the European XFEL.

Based on these results, there are still a number of directions which should be further investigated in the future:
First, the parameters under consideration corresponded to soft and hard X-ray pulses whereas the promising ELI facility will operate in the optical regime.
Accordingly, one should try to improve on the numerics so that one could reliably calculate the particle production in combinations of optical and X-ray pulses.

Secondly, the potential of optimization problems and techniques has hardly been exhausted:
In this work, we have employed an optimization algorithm which is based on a simple multi-start method.
However, there are various alternative algorithms to perform global optimization such as metaheuristic algorithms which could be more efficient for larger parameter spaces.
On the other hand, the optimization does not necessarily have to regard the particle number:
For instance, the requirement of specific spectral properties could in principle be formulated as an inverse scattering problem.

Finally, the long-time goal is to perform an ab initio simulation of particle production in a realistic laser collision in space and time, going beyond the locally-constant field approximation or the assumption of spatial homogeneity.
To this end, it is necessary to take into account the spatial inhomogeneity as well as magnetic fields.
In principle, approaches like the Dirac-Heisenberg-Wigner formalism \cite{BialynickiBirula:1991tx,Hebenstreit:2011wk}, worldline instantons \cite{Gies:2005bz,Dunne:2005sx} or real-time lattice gauge theory \cite{Hebenstreit:2013qxa} can deal with this problem, however, numerical simulations for realistic parameters are computatinally not feasible yet.

\subsection*{Acknowledgments}

We thank A.~D.~Bandrauk, G.~V.~Dunne, R.~Grobe, M.~Marklund and Q.~Su for fruitful discussions during the KITP program 'Frontiers of intense laser physics' as well as R.~Alkofer, C.~Kohlf\"urst, E.~Lorin, M.~Mitter and G.~von~Winckel for collaboration on related work. F.~Fillion-Gourdeau thanks S.~MacLean for many discussions and constant support. F.~Hebenstreit acknowledges support from the Alexander-von-Humboldt Foundation. This research was supported in part by the National Science Foundation under Grant No. NSF PHY11-25915.

\end{document}